\documentclass[twoside,twocolumn]{article}

\usepackage{blindtext} 

\usepackage[sc]{mathpazo} 
\usepackage[T1]{fontenc} 
\usepackage{microtype} 

\usepackage[english]{babel} 

\usepackage[hmarginratio=1:1,top=32mm,columnsep=20pt]{geometry} 
\usepackage[hang, small,labelfont=bf,up,textfont=it,up]{caption} 
\usepackage{booktabs} 
\usepackage{amsmath} 
\usepackage{graphicx}

\usepackage{lettrine} 

\usepackage{enumitem} 
\setlist[itemize]{noitemsep} 

\usepackage{abstract} 

\usepackage{titlesec} 
\renewcommand\thesection{\Roman{section}} 
\renewcommand\thesubsection{\roman{subsection}} 
\titleformat{\section}[block]{\large\scshape\centering}{\thesection.}{1em}{} 
\titleformat{\subsection}[block]{\large}{\thesubsection.}{1em}{} 

\usepackage{fancyhdr} 
\usepackage{titling} 

\usepackage{hyperref} 

\pretitle{\begin{center}\Huge\bfseries} 
\posttitle{\end{center}} 
\title{Low Rank Non-Negative Matrix Factorization with D-Wave 2000Q} 
\author{%
\textsc{Daniele Ottaviani}\thanks{CINECA, via Magnanelli 6/3, 40033 Casalecchio di Reno, Bologna (Italy).} \\[1ex] 
\normalsize CINECA \\ 
\normalsize \href{mailto:d.ottaviani@cineca.it}{d.ottaviani@cineca.it} 
\and 
\textsc{Alfonso Amendola}\thanks{ENI S.p.A., via Felice Maritano 26, 20097 San Donato Milanese, Milano (Italy)} \\[1ex] 
\normalsize ENI S.p.A. \\ 
\normalsize \href{mailto:alfonso.amendola@eni.com}{alfonso.amendola@eni.com} 
}
\date{\today} 
\renewcommand{\maketitlehookd}{%
\begin{abstract}
\noindent In this article we want to demonstrate the effectiveness of the new D-Wave quantum annealer, \texttt{D-Wave 2000Q}, in dealing with real world problems. In particular, it is shown how the quantum annealing process is able to find global optima even in the case of problems that do not directly involve binary variables. The problem addressed in this work is the following: taking a matrix $V$, find two matrices $W$ and $H$ such that the norm between $V$ and the matrix product $W\cdot H$ is as small as possible. The work is inspired by O'Malley's article \cite{article:omalley}, where the author proposed an algorithm to solve a problem very similar to ours, where however the matrix $H$ was formed by only binary variables. In our case neither of the two matrices $W$ or $H$ is a binary matrix. In particular, the factorization foresees that the matrix $W$ is composed of real numbers between 0 and 1 and that the sum of its rows is equal to 1. The QUBO problem associated with this type of factorization generates a potential composed of many local minima. We show that simple forward-annealing techniques are not sufficient to solve the problem. The new D-Wave 2000Q has introduced new solution refinement techniques, including reverse-annealing. Reverse-annealing allows to explore the configuration space starting from a point chosen by the user, for example a local minimum obtained with a precedent forward-annealing. In this article we propose an algorithm based on the reverse annealing technique (that we called adaptive reverse annealing) able to reach global minimum even in the case of QUBO problems where the classic forward annealing, or uncontrolled reverse annealing, can not reach satisfactory solutions.
\end{abstract}
}

\begin{document}

\maketitle


\section{Introduction}

\lettrine[nindent=0em,lines=3]{T} he problem of non-negative matrix factorization (NMF) is a problem of great interest for companies and research institutes. There are a lot of classic algorithms for factorizing a matrix in the product of two non-negative matrices, but in many cases none of them are able to reach global optima. So far the D-Wave quantum annealer has been used to solve many kind of problems, including real-world applications. All these problems, however, have in common the fact that they depend directly on binary variables. In his article \cite{article:omalley}, O'Malley showed how D-Wave is able to solve non-negative matrix factorization problems where one of the two matrices is composed only of binary variables (NBMF). In this article, we show how, with advanced techniques like the \emph{reverse annealing}, D-Wave is able to solve more generic NMF problems, where both matrices resulting from factoring are composed of real numbers. The approach chosen for NMF is the classic method of the Alternating Least Squares (ALS): choosing an initial matrix $H$, we search for a matrix $W$ such that the Frobenious norm 
\begin{equation}\label{eq:norm}
||V-W\cdot H||_F
\end{equation}
is minimal. Subsequently, obtained the matrix $W$, a new matrix $H$ is sought such that the same norm is minimal. By repeating this process for a certain number of iterations, two matrices $W$ and $H$ are obtained such that the norm described in equation \eqref{eq:norm} appears to be as small as possible.

The problem of our interest has very specific characteristics. In particular, it is required to factorize a given matrix $V\in \mathbb{R}^{n\times m}$ into two matrices $W\in \mathbb{R}^{n \times k}$ and $H \in \mathbb{R}^{k \times m}$, where $k=3$, with the matrix $W$ composed of elements between $0$ and $1$ ($W_{ij}\in [0,1]$) and such that the sum of its rows is always equal to $1$. No particular property is required for the matrix $H$.

The work is organized as follows: in Section 2, the decomposition of the minimization problem and the construction of the QUBO problem associated with the factorization are explained. Also, we explain the technique of \emph{adaptive reverse annealing} used to refine the solutions obtained by forward annealing. In section 3, the results obtained so far are highlighted. In particular, it is shown how it is possible to achieve global optimum results by refining the solutions found through cycles of forward annealing with cycles of reverse annealing, gradually increasing the area of research around the point of minimum obtained. In the last section we express considerations on the work done so far and on possible future applications.

\section{Methods}
In this section we present the decomposition of the original problem, necessary to implement the algorithm with the D-Wave 2000Q, the formulation of the QUBO problem and the description of the technique of \emph{adaptive reverse annealing} used to refine the solutions obtained by forward annealing.
\subsection{Problem Decomposition}
The number of variables that we can use for solving the problem is limited, so we have to decompose the NMF problem. The minimization of the norm described in equation \eqref{eq:norm} can be decomposed in this way:
\begin{equation}\label{eq:dec}
V=W\cdot H \Rightarrow \begin{cases} 
      V_j = W\cdot H_j\\
      V_i = H^T\cdot W_i 
   \end{cases}
\end{equation}

where $V_j$ is the $j$-th column of the matrix $V$, $H_j$ is the $j$-th column of the matrix $H$, $V_i$ is the $i$-th row of the matrix $V$ and $W_i$ is the $i$-th row of the matrix $W$.

Starting from an initial matrix $H$, the matrix W is searched one row at a time solving 

\begin{equation}\label{eq:wnorm}
min_{W_i}||V_i-H^T\cdot W_i||_2
\end{equation}

and the matrix H is searched one column at a time solving

\begin{equation}\label{eq:hnorm}
min_{H_j}||V_j-W\cdot H_j||_2
\end{equation}

The minimization of these norms occurs normally considering the squares of the equations \eqref{eq:wnorm} and \eqref{eq:hnorm}. Moreover, as regards the equation \eqref{eq:wnorm}, it is good to add a constraint term that maintains the sum of the rows of the $W$ matrix equal to $1$. Consequently, the functionals to be minimized are:

\begin{equation}\label{eq:wnorm2}
min_{W_i}\left( ||V_i-H^T\cdot W_i||_2^2 + \left( 1 - \sum_j W_{ij} \right)^2\right)
\end{equation}

\begin{equation}\label{eq:hnorm2}
min_{H_j}||V_j-W\cdot H_j||_2^2
\end{equation}

\subsection{QUBO problem}

The generic QUBO problem has the following form:

\begin{equation}
\sum_e a(e)q_e + \sum_{e<f} b(e,f)q_eq_f
\end{equation}

where $a(e)$ are the \emph{linear} coefficients and $b(e,f)$ are the \emph{quadratic} coefficients of the problem.

The first thing we need to do before writing the QUBO problem related to our problem is to find an appropriate form to represent the numbers we need using only binary variables. In this context we will limit ourselves to considering the elements of the matrix $W$, which have very specific restrictions ($W_{ij} \in [0,1]$, $\sum_j W_{ij} =1$).
A generic element $W_{ij}$ is written in this form:
\begin{equation}\label{eq:binw}
W_{ij} = \mathbf{c}\cdot\sum_{k=0}^N 2^{k}q_k
\end{equation}
with $\mathbf{c}=0.001$ and $N=9$. The number of binary variables used to represent a single number is $N+1$ (so, in our case, every element of $W$ is represented using $10$ binary variables). In this way, we can represent every number $W_{ij} \in [0,1.023]$, with 3 significant figures.

After that, knowing that every row of matrix $W$ has $k$ elements, we define
\begin{equation}\label{eq:d}
D_{je}^i =\begin{cases} 
      2^{e - j(N+1)} \cdot c & \mbox{if } e\in K\\
      0 & \mbox{if } e\notin K
   \end{cases}
\end{equation}
where the set $K\equiv \{j(N+1),\ldots,j(N+1) + N\}$, $N=9$ and $c=0.001$. Here, $j\in\{0,\ldots,k\}$ is the column index of the matrix $W$. So, if $j=0$, the first 10 qubits are considered (the set $K$ becomes $\{0,\ldots,9\}$), if $j=1$ we consider the second 10 qubits ($K\equiv \{10,\ldots,19\}$) and so on. Index $i$ is the row index of the matrix $W$.
With this notation we can represent the $k$ elements of a generic row of matrix $W$ using a single vector of $k(N+1)$ qubits.

The coefficients for the QUBO problem related to equation \eqref{eq:wnorm2} are:

\begin{equation*}
\begin{split}
a_i(e)&= \sum_j \bar{H}_j^2 \left(D_{je}^i\right)^2 + 2\sum_{j<l} \left(HH^T\right)_{jl}D_{je}^iD_{le}^i \\
&-2 \sum_j \left(VH^T\right)_{ij}D_{je}^i+\sum_j \left(D_{je}^i\right)^2 \\
&-2 \sum_j D_{je}^i + 2\sum_{j<l} D_{je}^iD_{le}^i
\end{split}
\end{equation*}
\begin{equation*}
\begin{split}
b_i(e,f)&= 2\sum_j\bar{H}_j^2D_{je}^i D_{jf}^i + 2\sum_{j<l} \left(HH^T\right)_{jl} D_{je}^i D_{lf}^i\\
&+2\sum_j D_{je}^i D_{jf}^i + 2\sum_{j<l} D_{je}^i D_{lf}^i 
\end{split}
\end{equation*}

where
\begin{itemize}
\item $\bar{V}_j^2 = \sum_k V^2_{kj}$
\item $\bar{H}_j^2 = \sum_k H^2_{kj}$
\item $\left(HH^T\right)_{jl} = \sum_k H_{jk}H_{lk}$
\item $\left(VH^T\right)_{ij} = \sum_k V_{ik}H_{jk}$
\end{itemize}

The coefficients depend on $i$ because they change in value according to the row that is being analyzed.
\subsection{Embedding the problem}
\begin{figure}
\includegraphics[width=1\linewidth]{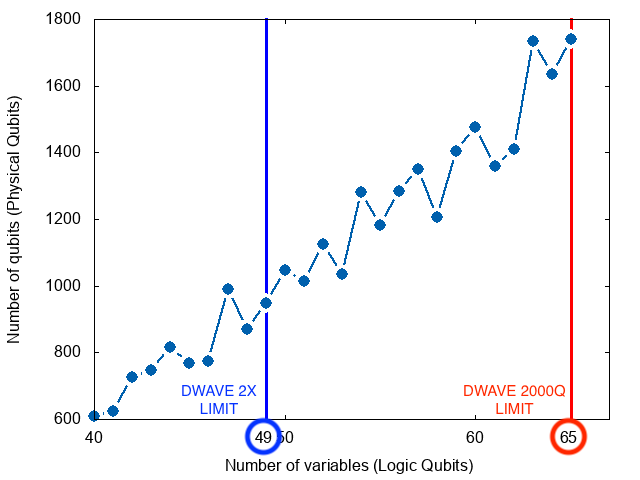}
\caption{The figure shows the relationship between physical qubits and logical qubits in the case of a QUBO problem whose graph is completely connected (i.e. $b_i(e,f) \neq 0$ $\forall i,e,f$). The problem faced in this work generically falls into this category. As we can see from the figure, the maximum number of logical variables usable with D-WAVE 2000Q for a problem like ours is 65 (blue benchmark in the figure). The previous model was able to embed 49. Conditions of sparseness of the matrix to be factored can increase this limit.}
\label{fig:full}
\end{figure}

As we can see in Figure \ref{fig:full}, the number of variables needed to simulate a fully connected graph is much lower than the number of physical qubits that make up the QPU of the D-WAVE 2000Q. This is due to the fact that the Chimera graph, the network of connections that unites the qubits of the D-Wave annealer, is not a complete connected graph. In case you want to solve a QUBO problem whose coefficients imply a complete graph, you must join several qubits in order to make them act as if they were a single vertex of the graph. This procedure is known as \emph{embedding}. In our case, the maximum number of logical variables usable on the D-WAVE 2000Q is 65. Since each number is expressed as a function of 10 qubits, the maximum number of variables usable in a single problem is 6. Although this seems to be a heavy restriction , we note that in the problem chosen to be addressed in this article, the number of variables required (ie the rank of the matrix factorization) is 3. Consequently, the limitation posed by D-WAVE 2000Q on the maximum number of variables usable in the same problem does not no restriction on the resolution of the problem itself.

\subsection{Reverse annealing}

The use of the reverse annealing technique to refine the solutions obtained with the simple forward annealing is a central point for the job. The coefficients of the QUBO problem calculated in the previous section differ from each other by several orders of magnitude. This is mainly due to the fact that the elements of the matrices to be factored are not binary variables but are real numbers expressed on a binary basis (each qubit is associated with a different power of $2$).

This difference between orders of magnitude leads to the creation of many local minimum points, some very close to the global minimum, but still unsatisfactory for the purposes of our work. As we will see in the results section, simple forward annealing is not sufficient to achieve good results, even using the maximum number of cycles made available by the D-Wave machine. One way to reach global minima is to use the \emph{reverse annealing} technique. Reverse annealing, one of the new features made available by the last version of the quantum annealer, D-Wave 2000Q, allows to perform annealing cycles in search of global minima starting from a configuration chosen by the user. Normally, forward annealing always starts from the same initial configuration. Until the D-Wave 2000Q, it was not possible to choose this configuration.

Reverse annealing allows the user to manipulate different parameters in order to obtain the best solutions. The parameter that we will take most into consideration in this work is the \emph{holding time} parameter, that is the time during which the annealing cycle seeks better minima around the starting point. As fully explained in the article \cite{article:rev}, the breadth of the local search is related to the \emph{reversal distance}, which specifies how far we anneal backward.

With the term \emph{reversal distance} we mean the average of the \emph{Hamming distance} between consecutive solutions of the same reverse annealing. The Hamming distance, in turn, is a distance that applies to strings of characters (and therefore also binary variables): it is calculated taking into account the differences between elements placed in the same position. It is possible to correlate the reversal distance of the solutions with the \emph{holding time} parameter of the reverse annealing. The holding time is directly related to the search space of reverse annealing cycles: the longer this time, the greater the space of the configurations analyzed. Still in the same article, it is explained that the reversal distance between the solutions must be chosen with care: a low value is equivalent to search in a space very close to the starting point. This could lead to the failure of reverse annealing, returning as a result the same initial point given as input.

On the contrary, instead, a too big reversal distance could lead the system to forget the starting point, returning as a result a configuration that corresponds to a point of minimum even worse than the initial point. There is therefore a certain optimal band, within which the probability of finding a new ground state is maximized. In our work, we control the reversal distance with the holding time parameter, which establishes the maximum duration of each reverse annealing cycle. The higher the value of the holding time parameter, the greater the value of the calculated reversal distance.

Calculating this band is not very simple: at the state of the art it must be sought through a study of the parameters in relation to the problem to be solved. The method chosen in this work to look for the right holding time for each reverse annealing cycle is an adaptive method: the holding time parameter, initially set at low values, is gradually increased until a new ground state is obtained (or up to a predetermined time limit).

In the next section we will see a first application of the adaptive reverse annealing method in the resolution of a linear system, a solution sought by minimizing the norm presented in equation \eqref{eq:wnorm2}.


\section{Results}

\subsection{Solving a Linear System}

The first problem we have chosen to solve was the resolution of a linear system, whose vector of unknowns follows the restrictions imposed on the rows of the $W$ matrix of our original factorization problem ($W_{ij} \in [0,1]$, $\sum_jh W_{ij}=1$). In particular, we will show how it is possible to solve this linear system by minimizing the proposed norm in equation \eqref{eq:wnorm2}. Solving a linear system whose vector of unknowns follows the rules described by equation \eqref{eq:wnorm2} corresponds to carrying out an iteration of the matrix factorization knowing in advance the point of global minimum to be reached.

The chosen linear system is:
\begin{equation}
H^T\cdot w = V
\end{equation}
where
\[
H=\begin{bmatrix}
    1.301 & 0.440 & 0.672 & 0.218 & 0.024\\
   0.125 & 0.342 & 0.709 & 0.427 & 0.036\\
   0.187 & 0.082 & 0.802 & 0.520 & 0.038
\end{bmatrix}
\]
\[
V=\begin{bmatrix}
0.365 & 0.232 & 0.748 & 0.435 & 0.035
\end{bmatrix}
\]
The correct solution is

\[w=\begin{bmatrix}
    0.178 & 0.333 & 0.489
\end{bmatrix}
\]
\paragraph{Solving with forward annealing}

\begin{figure}
\includegraphics[width=1\linewidth]{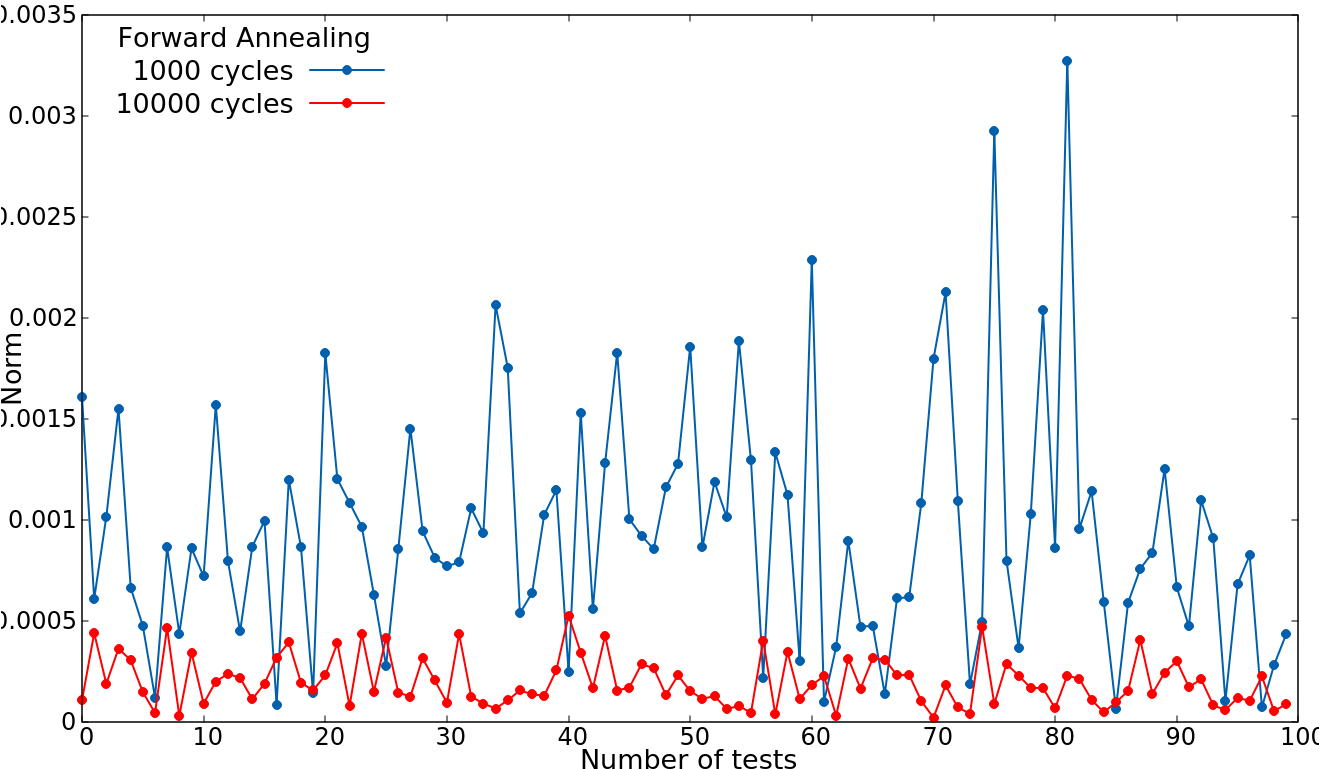}
\caption{Results obtained through forward annealing. In blue we can see the results of 100 tests performed each with 1000 cycles of annealing. In red instead we can observe the results of 100 tests carried out each with 10000 annealing cycles. Although very close to the optimal solution, none of these tests has ever reached the global minimum (represented, in this case, by the analytical solution of the linear system)}
\label{fig:100_1000}
\end{figure}

\begin{figure}
\includegraphics[width=1\linewidth]{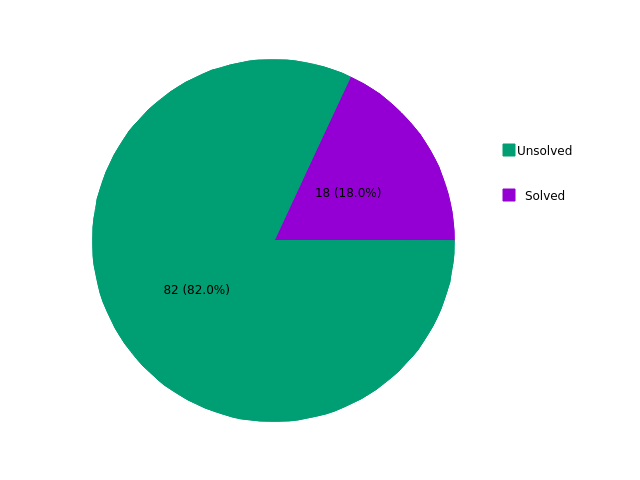}
\caption{Pie chart representing the results obtained following the reverse annealing algorithm proposed in the paragraph \emph{Solving with forward annealing plus reverse annealing}. Of the 100 tests carried out, 18 have reached the global minimum. The others were trapped in some local minimum.}
\label{fig:h_no_rev}
\end{figure}

First of all, we looked for solutions of the linear system using different cycles of forward annealing. Figure \ref{fig:100_1000} shows the results of 200 tests, 100 performed using 1000 annealing cycles (blue points in the figure) and 100 performed using 10000 annealing cycles (red points in the figure). The results, especially those obtained with 10000 annealing cycles, are very close to the analytical solution of the system. Yet, no forward annealing cycle calculate the exact solution. The QUBO problem we want to solve presents not inconsiderable differences between the orders of magnitude of its coefficients. This fact generates a large number of local minimum points in the potential associated with the problem. Evidently, each forward annealing seems to fall into one of these local minimum points.
\paragraph{Solving with forward annealing plus reverse annealing}
The D-Wave 2000Q presents, unlike previous models, a series of new features, which allow the user to manage some aspects of the annealing cycles. One of these is the possibility to perform a \emph{reverse annealing}. Unlike forward annealing, reverse annealing allows the user to choose the starting point of the search for the minimum. It allows therefore to search in the surroundings of a given point a configuration that leads to lower energy minima. We have tried to refine the solutions obtained with forward annealing using reverse annealing. In particular, the strategy adopted at this stage was:
\begin{enumerate}
\item Search for a \emph{good} local minimum point with forward annealing
\item Perform a reverse annealing starting from the result obtained in point 1. Is the obtained result the global minimum? If yes, stop. Else, repeat the reverse annealing, using this result as new starting point.
\end{enumerate}
With this method we have begun to observe some results: out of 100 tests carried out following the algorithm described above, 18 have reached the global optimum (see figure \ref{fig:h_no_rev}). The others, unfortunately, have continued to stop on local minima, failing to get out of it. The algorithm just presented has a threshold of reverse annealing attempts. If this threshold is exceeded, the problem is considered unsolved.

\paragraph{Solving with forward annealing plus reverse annealing with adaptive strategy}
After seeing that the reverse annealing technique can be successfully used to reach the global minimum point for our QUBO probelam associated with the resolution of a linear system, we have begun to think of a way to improve its performance. The introduction of reverse annealing has allowed us to get the right result 18 times out of 100. That is better than the result obtained using the simple forward annealing, but it is not yet enough.

\begin{figure}
\includegraphics[width=1\linewidth]{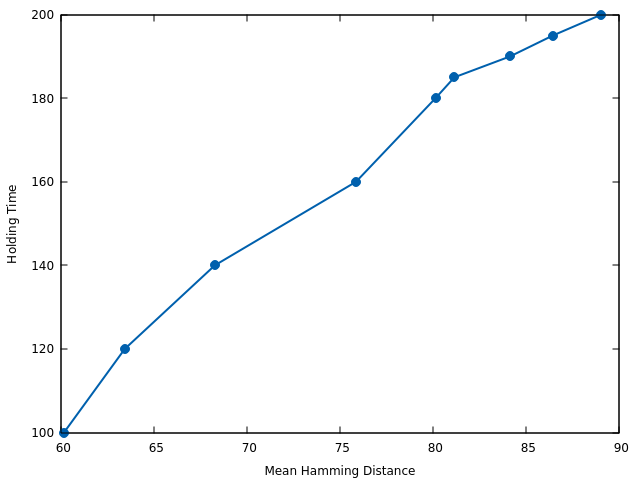}
\caption{Correlation between mean hamming distance and holding time. The data were extrapolated from one of the 100 tests carried out in an attempt to solve the linear system with an adaptive reverse annealing strategy}
\label{fig:corr_ht}
\end{figure}
As shown in article \cite{article:rev}, the probability of obtaining a new ground state through the reverse annealing technique can be influenced, and therefore maximized. In article \cite{article:rev}, the authors show a strong correlation between the Hamming distances calculated between solutions derived from consecutive reverse annealing cycles and the probability of finding a new ground state. They showed, in particular, that there are two threshold values for the average of the Hamming distances between consecutive solutions such as to influence the success of the reverse annealing search. If the average Hamming distance falls below the minimum threshold, reverse annealing is likely to fall back to the same starting point. If, on the other hand, the average Hamming distance falls above the maximum threshold value, the search moves away too much from the starting point, forgetting the associated energy value, and will probably end up at a local minimum point energetically higher than the starting point. Therefore, in order to maximize the probability of a successful search with reverse annealing, it is necessary to analyze \textit{a priori} the average of the Hamming distances between consecutive solutions generated by different searches.
During our research, we figured out that the \texttt{holding time} parameter of the reverse annealing search is strongly correlated with the mean of the Hamming distances calculated on the solutions obtained (see figure \ref{fig:corr_ht}). The correlation is direct: the more you ask the annealing to wait, the more the search moves away from the starting point, the more the average of the Hamming distances increases. Considering that, we have opted for an adaptive approach. The term \emph{adaptive}, in this case, means that the algorithm looks for the right value of the mean of Hamming distances by adapting the \texttt{holding time} parameter of the reverse annealing search, having observed the direct correlation. Starting from low values of holding time, we observe the final solution of the associated annealing search: if the search ends at the starting point, we repeat it by increasing the holding time parameter, thus increasing the mean of the Hamming distances of the solutions. Schematically, the new algorithm consists of the following steps:

\begin{enumerate}
\item Search a good local minimum point with forward annealing
\item Perform a reverse annealing starting from the result obtained in point 1. Is the obtained result the global minimum? If yes, stop. Else, repeat the reverse annealing, using this result as new starting point.
\item If the result is the same as the starting point, repeat the reverse annealing using the same starting point but increasing the holding time until you find a new ground
state (or until the maximum number of attempts is reached). Once found, reset the holding time parameter to its default value.
\end{enumerate}

\begin{figure}
\includegraphics[width=1\linewidth]{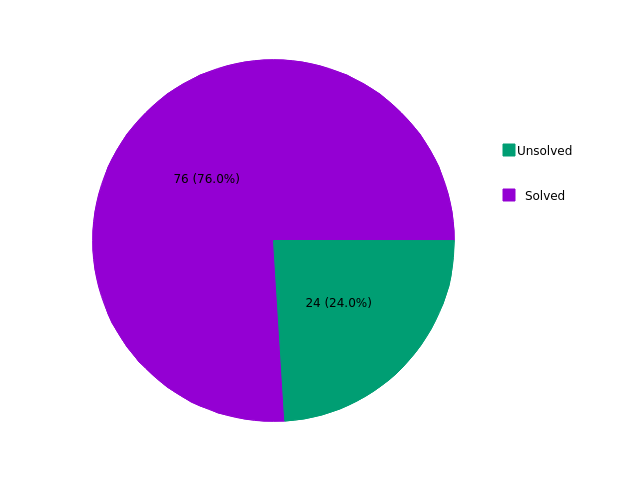}
\caption{Pie chart representing the results obtained following the reverse annealing algorithm proposed in the paragraph \emph{Solving with forward annealing plus reverse annealing with adaptive strategy}. Of the 100 tests carried out, 76 have reached the global minimum.}
\label{fig:h_rev}
\end{figure}

In this way, we do our research with different reverse annealing around a local minimum point, gradually expanding the search range, paying attention to increase the waiting time in a sufficiently gradual manner, so as to maximize the probability of falling into the region of \emph{maximum probability of finding a new ground state} described in article \cite{article:rev}. In fact, as shown in chart \ref{fig:h_rev}, out of 100 tests performed with this new algorithm, 76 are successful. The gradual increase of the holding time has made it possible to prevent the annealing from stopping at a single local minimum point (as was the case with reverse annealing carried out without any particular precautions). However, even using this strategy about a quarter of the attempts at solving the linear system did not succeed. This is because in some cases it was not possible to reach the global minimum within a maximum of 50 attempts, the upper limit that was chosen to avoid infinite loops.

Next, we tried to apply this new algorithm to the factorization of a small matrix.

\subsection{Non-negative Matrix Factorization}

At this stage of the project, we decided to test the adaptive search algorithm on the factorization of a small matrix. In particular, we have chosen a matrix $V\in \mathbb{R}^{2\times 2}$ and we have solved
the factoring problem using an ALS technique both on classical computer and on D-Wave. For the classic computer implementation, we used the python library \texttt{lsqnonneg.py}, a python library written to solve minimization problems like those described by the equations \eqref{eq:wnorm2} and \eqref{eq:hnorm2}.

We decided to try our algorithm with a small $2\times 2$ matrix for reasons relating exclusively to our computational resources. In theory, the algorithm is able to factorize any matrix $V\in \mathbb{R}^{n\times m}$ $\forall n,m\in \mathbb{N}$ into the product of two matrices $W\in \mathbb{R}^{n\times k}$ and $H\in \mathbb{R}^{k\times m}$. The only limitation of our algorithm is in the maximum rank of the factorization, that is the variable k relative to the dimensions of the two factoring matrices. This dimension, we must remember, must be less than or equal to 6. Although this restriction seems to be a strong limitation, in practice it is not, since this type of factoring often requires explicitly that the rank is kept very low.

For the implementation on D-Wave, however, we used a mixed approach: for the calculation of the $H$ matrix we used the same python library used for the classical implementation;
for the calculation of the $W$ matrix, instead, we used our algorithm composed of forward annealing plus an adaptive search with reverse annealing. For a better comparison of the results, we used the same initial matrices in both algorithms.

\begin{figure}
\includegraphics[width=1.1\linewidth]{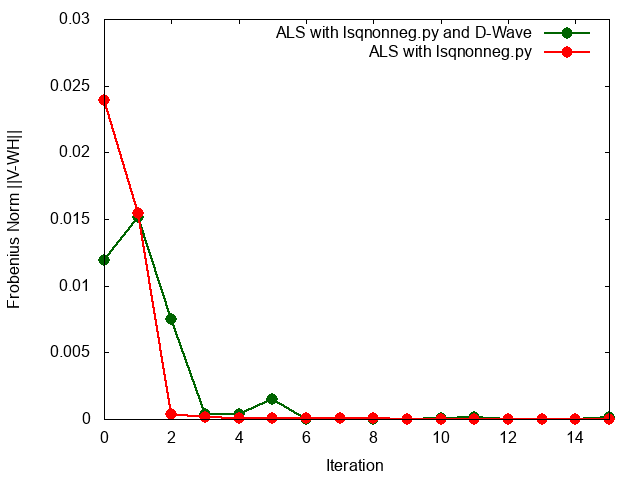}
\includegraphics[width=1.1\linewidth]{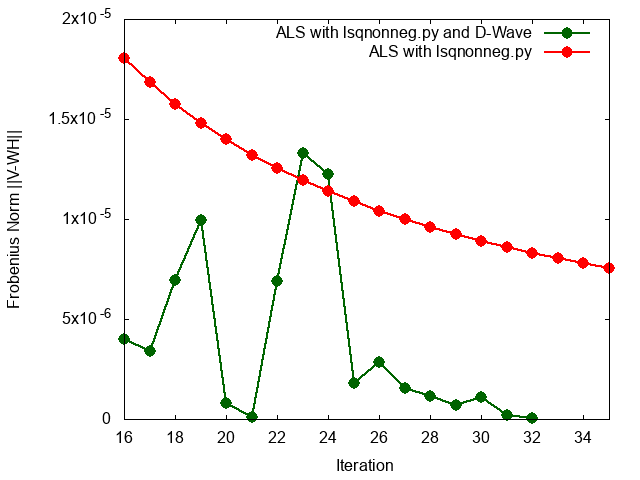}
\caption{Comparison between factorization with classical method (red line) and with \emph{mixed} D-Wave method (green line). The graph represents the trend of the norm $||V-W\cdot H||_F$, as the number of iteration increases. Both algorithms received the same initial $H$ matrix as input, randomly calculated. The first image represents the convergence of the algorithm in the first 15 iterations. The second image represents iterations from 16 to 35 (with a different zoom). The mixed method achieved the best result in 32 iterations, while the classical method reaches the limit of 10000 iterations without achieving a result of the same order of magnitude. In particular, further calculations show that the classical method achieved the best result after about 67000 iterations, without ever reaching a factorization with $||V-W\cdot H||_F$ of the same order of that obtained from the mixed method.}
\label{fig:factorization}
\end{figure}

Figure \ref{fig:factorization} shows an example of comparison between the performance of the two methods. In this particular case, factorization with the aid of D-Wave produced better results than its classical counterpart. In fact, within 50 iterations (maximum limit for the D-Wave based method, due to computational time requirements) the classic-D-Wave method obtained a factorization with a $||V-W\cdot H||_F$ norm equals to $5.98835340634691e-08$, while the classic method after 10000 iterations stopped at a norm equals to $2.25212726195e-07$ (and it achieved the best result after about 67000 iterations, with a norm equal to $1.50292153373e-07$).

As we can see from figure \ref{fig:factorization}, unlike the classic ALS method fully implemented by the python library \texttt{lsqnonneg.py}, the classic-D-Wave ALS method does not converge uniformly towards the solution. In some iterations, in fact, the $||V-W\cdot H||_F$ norm grows slightly instead of decreasing. The general trend is however descending. These reversal points correspond to wrong minimizations, ie minimizations where the adaptive reverse annealing method failed to find a satisfactory minimum (it is then stuck in some local minimum). These points, however, do not involve any problem for the convergence of the method.

\subsection{QPU Computational Time}

The purpose of this section is to provide an adequate estimate to understand the computational times related to the minimization operations performed with the D-Wave 2000Q.
Estimates for calculating computational times strictly related to the use of D-Wave (\emph{QPU computational times}) only concern the times of annealing cycles, modified by the number of calls and reverse annealing techniques.

D-Wave 2000Q, the latest model of the quantum annealer produced by D-Wave, is able to complete an annealing cycle in just 1 $\mu s$. Normally, each minimization process consists of a number of annealing cycles to find the solution. In our case, the forward annealing process has always involved the maximum number of cycles available, ie 10000. Consequently, the QPU-time required to complete a single minimization with a forward annealing is equal to 10000 $\mu s$, or 0.01 seconds.

It is very interesting to note that these computational times do not depend on the size of the system: the only limitation on the number of variables that can be used to solve a problem is given by technological limitations, that is the number of qubits and couplers of the annealer. In principle it is possible to solve minimization problems with the same computational time regardless of their size. What will affect the time for the solution will be only the number of cycles required.

In our work, after looking for an optimal solution with 10000 cycles of forward annealing, we decided to refine the solution using a series of reverse annealing calls, each in turn co-imposed by 10000 cycles. Reverse annealing is more expensive than forward annealing: it provides an \texttt{holding time} parameter, expressed in microseconds, during which the algorithm looks for better solutions near the starting point. As we have already explained in the course of the work, the strategy chosen to efficiently apply reverse annealing is an adaptive strategy: the holding time is gradually increased until a new ground state is reached (or until the maximum of allowed iterations is reached). The maximum number of iterations allowed in our case is 50, while the maximum value chosen for the holding time parameter is 200 $\mu s$.

In general, the additional QPU-time due to reverse annealing calls can be calculated with a simple formula:
\begin{equation}\label{eq:trev}
T_{rev}=N_{calls}\times (T_h + 1) \times N_{cycles}\quad \mu s,
\end{equation}
where $N_{calls}$ is the number of calls needed to get a satisfying ground state (maximum 50), $T_h$ is the value of the holding time parameter expressed in microseconds (maximum 200) e $N_{cycles}$ it is the number of cycles necessary to complete a single call of reverse annealing (10000, in our work). Considering the worst case, that is $N_{calls}=50$ and $T_h=200$, the QPU-time needed to arrive at a satisfactory solution increases by $50 \times 201 \times 10000 = 100500000 \mu s = 100.5 s$.

Unfortunately, the technological limitations of D-Wave do not allow dealing with a factoring problem with a single annealer call. Instead, it is necessary to decompose the problem. In our case, we decided to tackle the decomposition by minimizing one line at a time (if we think of the matrix W. One column at a time if we think of the matrix V). Furthermore, it is necessary to repeat the procedure for a certain number of iterations. In this article we have decided to proceed with a maximum of 50 iterations. In any case, the computational times increase considerably: they increase according to the rule:

\begin{equation}\label{eq:tfac}
T_{factorization} = N_{iterations} \times N_{rows} \times T_{rev},
\end{equation}

where $T_{rev}$ is the time calculated in equation \eqref{eq:trev}, $N_{iterations}$ is the number of iterations necessary for the factorization and $N_{rows}$ is the number of rows (or column) for the matrix to be factorized. If we calculate this time taking into account the factoring carried out in the previous section, always considering the worst case, we get $50 \times 2 \times 100.5 = 10050.0 s = 2.79 h$. This computational time, which only concerns the use of the quantum annealer QPU, may seem very high. In fact it is, but we must also take into account many other factors when we analyze it. As already mentioned, this computational time does not depend on the size of the system, except for technological limitations: very large systems can be managed much more efficiently using the D-Wave. Moreover, the technological evolution of D-Wave has so far significantly reduced the times of a single annealing cycle: from 20 $\mu s$ of the model with 512 qubits we have passed to 1 $\mu s$ of the current model. Any further decrease in these annealing times will inevitably lead to improved performance.


\section{Conclusions}
In this article we have shown that the D-Wave 2000Q, built to solve QUBO problems, can also be used to solve QUBO problems involving real numbers written on a binary basis. As far as we know, it is the first time that someone tries to solve a problem not based directly on binary variables with a quantum annealer. In the literature, to date, there are no similar works.

Normally, these kinds of problems are difficult to resolve through annealing techniques, be it quantum or simulated. The reason lies mainly in their construction: the large difference between the orders of magnitude of the linear and quadratic coefficients associated with each qubit causes a very large number of local minimum points to be generated in the potential associated with the problem. Points where minimal research techniques based on principles of annealing tend to get stuck.

In this paper we show three different progressive approaches to try to solve a problem of minimization linked to the well-known problem of the Nonnegative Matrix Factorization (NMF). Starting from the simple forward annealing, with which we were not able to have satisfactory results, we have gradually integrated the solutions supplied with other refining techniques proposed by D-Wave's new quantum annealer, first of all the reverse annealing technique.

We have noticed, in fact, that if we take the results obtained with the forward annealing as starting points for subsequent cycles of reverse annealing, we can reach the global minimum point. In particular, we proposed an adaptive reverse annealing technique, able to gradually adapt the search space of the solutions around a local minimum point. With this technique we have managed to reach the global minimum point 75 times out of 100, against the 0/100 obtained using forward annealing and 18/100 obtained using forward annealing and reverse annealing without any control over its parameters.

We have demonstrated the effectiveness of the proposed technique either by solving a linear system and by factoring a small matrix $V$ in the product of two matrices $W$ and $H$, comparing the results obtained by our algorithm with the results obtained by a classical algorithm for NMF. The factoring through D-Wave 2000Q was done with a mixed approach: while the $W$ matrix was calculated using the quantum computer, the $H$ matrix was calculated using the classical algorithm. This choice has a motivation of interest: in our work we have given particular attention to the calculation of the $W$ matrix, which must be built respecting some properties. In particular, it must consist of elements between 0 and 1 and the sum of its rows must always be equal to 1.

Tests showed that it is possible to carry out an NMF also using the quantum annealer. In all the tests, the results of the two algorithms are very similar: both have always been able to reach a factorization characterized by a very small residual norm (i.e., the Frobenius norm of the difference of the original matrix $V$ and the dot product $W\cdot H$). In particular, in one of the tests carried out, the classical-quantum mixed method obtained a factorization characterized by a residual norm smaller than an order of magnitude with respect to the result obtained from the classical counterpart. It is also important to note that the mixed quantum-classical method requires much less iterations to reach significant factorizations: to obtain a good result, in fact, the algorithm that also involves the D-Wave 2000Q required only 50 iterations (approximately), while the classic algorithm required at least 10000.

These results are very encouraging: they show that the range of possible applications for D-Wave's quantum annealer does not stop at the problems that directly involve binary variables. The only limitations we have encountered in our work are to be attributed to the still limited number of variables that can be managed by the quantum annealer. The high QPU time required to complete a factorization can also be reduced by increasing the number of variables managed at the same time.
At the moment, in fact, the D-Wave 2000Q can manage 6 or 7 numerical variables at a time (considering that in our work the number of qubits used to represent a single numerical variable is 10). This happens because the connection graph related to our QUBO problem is a fully connected graph, and the embedding necessary to emulate a fully connected graph on the Chimera architecture of D-Wave 2000Q drastically reduces the number of qubits that can be used as logical variables, bringing it precisely at about 60. Under conditions of sparseness of the matrix of the quadratic terms associated with the QUBO problem, this number may increase.

If, as is logical to think, the models following the D-Wave 2000Q will be able to manage more variables than their predecessors, increasing both the number of qubits of the system and the number of connections between qubits, soon we can fully exploit the potential of the quantum annealer with every kind of problem.

\section{Acknowledgement}

Many thanks to Eni S.p.A. and CINECA. Without their support and interest, this breakthrough work would not have been possible.



\end{document}